\documentclass[10pt,conference]{IEEEtran}

\usepackage{algorithm}
\usepackage{algpseudocode}   
\usepackage{graphicx}
\usepackage{textcomp}
\usepackage{makecell}
\usepackage{subcaption}
\usepackage{listings}
\usepackage{fancyhdr}
\usepackage[dvipsnames]{xcolor}
\usepackage{multirow}
\usepackage{soul}
\definecolor{mygreen}{HTML}{2DA44E}
\usepackage{balance}
\usepackage{multicol}
\usepackage{xspace}
\usepackage{listings}
\usepackage{pifont}
\usepackage{xcolor}
\usepackage{threeparttable}
\usepackage{wrapfig}
\usepackage[numbers]{natbib}
\usepackage{tabularx}
\usepackage{tcolorbox}
\tcbuselibrary{listingsutf8}

\usepackage{hyperref}
\hypersetup{
  colorlinks,
  citecolor=blue,
  linkcolor=blue,
  urlcolor=blue,
}

\tcbset{
  colback=gray!10,
  colframe=black,
  listing only,
  listing options={basicstyle=\ttfamily\small,breaklines=true},
}

\usepackage{tcolorbox} 

\newcommand{\rqboxc}[1]{\begin{tcolorbox}[left=3pt,right=3pt,top=3pt,bottom=3pt,colback=gray!5,colframe=gray!40!black,before skip=5pt,after skip=5pt]#1\end{tcolorbox}}
\definecolor{brinkpink}{rgb}{0.98, 0.38, 0.5}

\newcommand{\phead}[1]{\vspace{1mm} \noindent {\bf #1}}

\newcommand{\tool}{\textit{TrajSpec}\xspace}
\newcommand{\baseline}{\textit{Agentic-Base}\xspace}

\usepackage{listings}

\usepackage{pmboxdraw, amsmath}

\usepackage{booktabs}
\definecolor{dkgreen}{rgb}{0,0.6,0}
\definecolor{gray}{rgb}{0.5,0.5,0.5}
\definecolor{mauve}{rgb}{0.58,0,0.82}
\definecolor{darkgreen}{rgb}{0.01, 0.75, 0.24}
\hypersetup{
  pdftitle={Bug Report Specification Refinement with Trajectory Guidance for Automated Program Repair},
  pdfauthor={S M Farah Al Fahim, Md Nakhla Rafi, Md Ahasanuzzaman, Zeyang Ma, Dong Jae Kim, Shaowei Wang, Tse-Hsun (Peter) Chen},
  pdfkeywords={Automated Program Repair, Bug Report Enhancement, Specification Refinement, Large Language Models, Repository-Level Repair, Software Maintenance}
}

\AtBeginDocument{%
  }

\begin{document}

\title{Bug Report Specification Refinement with Trajectory Guidance for Automated Program Repair}

\author{
\IEEEauthorblockN{
S M Farah Al Fahim\IEEEauthorrefmark{1},
Md Nakhla Rafi\IEEEauthorrefmark{1},
Md Ahasanuzzaman\IEEEauthorrefmark{1},
Zeyang Ma\IEEEauthorrefmark{1}\\
Dong Jae Kim\IEEEauthorrefmark{2},
Shaowei Wang\IEEEauthorrefmark{3},
Tse-Hsun (Peter) Chen\IEEEauthorrefmark{1}
}

\IEEEauthorblockA{
\IEEEauthorrefmark{1}
Software Performance, Analysis, and Reliability (SPEAR) Lab, Concordia University, Montreal, Canada\\
smfarahal.fahim@mail.concordia.ca,
mdnakhla.rafi@mail.concordia.ca,
m\_ahasa@live.concordia.ca\\
m\_zeyang@encs.concordia.ca,
peterc@encs.concordia.ca
}

\IEEEauthorblockA{
\IEEEauthorrefmark{2}
DePaul University, Chicago, USA \quad
dkim121@depaul.edu
}

\IEEEauthorblockA{
\IEEEauthorrefmark{3}
University of Manitoba, Winnipeg, Canada \quad
Shaowei.Wang@umanitoba.ca
}
}

\maketitle

\newcommand{\RQOne}{RQ1: How effective are \tool's refined bug reports at improving automated program repair performance?}
\newcommand{\RQTwo}{RQ2: Do \tool's refined reports improve repair performance across different downstream repair agents?}
\newcommand{\RQThree}{RQ3: How do the repository-based review and hierarchical evidence representation of \tool contribute to repair performance?}
\newcommand{\RQFour}{RQ4: What are the costs of generating \tool's refined bug reports and using them for automated program repair?}

\begin{abstract}
Bug reports serve as task specifications for repository-level automated program repair (APR) agents, but they often describe only the observed failure and omit repair-relevant information such as the failure-inducing behavior, behavioral requirement, and implementation scope. As a result, a repair agent may inspect irrelevant code, infer an incorrect requirement, or generate a patch that addresses the reported symptom without restoring the intended repository behavior. We present \tool, a trajectory-guided approach for repository-supported bug report specification refinement. Given an original report and a pre-fix repository, \tool runs a trajectory-collection agent and uses the resulting unverified trajectory as a source of trajectory-derived specification evidence. It organizes this evidence into a three-level representation consisting of a high-level interpretation of the issue, diagnostic findings supporting that interpretation, and concrete repository observations. \tool then generates a draft refined report and applies repository-based review to remove unsupported claims, revise uncertain claims, and add repository-supported details. We evaluate \tool on all 300 SWE-Bench Lite instances using Mini-SWE-Agent V2. \tool's refined reports improve Pass@1 from 41.00\% to 59.67\% with GPT-5-mini and from 54.67\% to 64.33\% with MiniMax M2.5. On a stratified sample of 100 instances, \tool's refined reports also improve Pass@1 from 41.00\% to 71.00\% with Agentless and from 47.00\% to 72.00\% with AutoCodeRover. Ablation results show that removing repository-based review or the hierarchical evidence representation reduces Pass@1 from 59.67\% to 48.00\% and 47.67\%, respectively. Overall, \tool provides actionable repository-supported context that consistently improves repair performance.
\end{abstract}

\section{Introduction}
\label{sec:introduction}

Large language models (LLMs) have enabled a new generation of repository-level automated program repair (APR) agents~\cite{yang2024swe,zhang2024autocoderover,bouzenia2025repairagent,zhang2026sgagent,tamoyan2026sherloc}. Given a bug report and a pre-fix repository snapshot, these agents search the repository, inspect source code, reason about the expected behavior, and generate patches that are validated against tests. Benchmarks such as SWE-Bench~\cite{jimenez2024swebench} have made this setting a central testbed for evaluating whether LLM-based agents can resolve real software issues from open-source repositories. In this setting, a bug report is not only a communication artifact for developers. It is the primary task specification from which the repair agent must infer what behavior is wrong, what behavior should hold, and which parts of the repository are relevant to the fix.

However, the information needed to describe the symptoms of a bug is not necessarily sufficient to guide repository-level repair. Prior work has identified report elements that support bug understanding and diagnosis, including reproduction steps, observed and expected behavior, affected components, stack traces, logs, and other diagnostic information~\cite{bettenburg2008makes,soltani2020significance,chen2021demystifying,chen2021pathidea,hirsch2020root,medeiros2024impact}. These elements help describe where and how a failure manifests, but they may not specify the repository-level behavior important for repair. In particular, the report may leave implicit the relevant implementation logic, the constraints that a correct fix must preserve, and the scope over which the expected behavior should hold. Human developers can recover such information through project knowledge, follow-up discussion, and repository exploration. A repair agent must instead infer it from the report and the pre-fix repository. When this inference is incomplete or incorrect, the agent may inspect irrelevant code, infer the wrong requirement, or generate a patch that removes the reported symptom without restoring the intended behavior~\cite{suri2026codescout,kuang2026reagent}.

Such incomplete or underspecified reports expose a specification problem in repository-level APR. Most recent work improves the repair agent through better search, localization, prompting, planning, or patch generation~\cite{xia2025agentless,zhang2024autocoderover,minisweagentV2,yang2024swe,yu2025orcaloca,xia2025live,wang2025openhands}. These improvements are important, but they still assume that the input report provides a sufficient task specification. 
Prior work has also explored report structuring, crash reproduction from stack traces, duplicate grouping, fault localization, and LLM-based report rewriting~\cite{rastkar2010summarizing,nayrolles2015jcharming,dang2012rebucket,acharya2025can}. These approaches make reports easier to read or connect reports to code artifacts, but they do not directly refine the report by gathering the repair specification needed by a downstream APR agent. The missing step is to expose the repair-relevant details that the report leaves implicit before downstream repair begins.

A trajectory-collection run can help recover some missing specification elements. When the trajectory-collection agent explores the repository for an underspecified report, its trajectory records the agent's search, code inspection, tool use, hypotheses, and revisions to its interpretation of the issue. Even without knowing whether any resulting candidate patch is correct, the trajectory can provide repository-supported evidence for refining the report, including relevant code locations, dependencies among affected components, and behavioral constraints that were not explicit in the original report.

However, raw trajectories are long, noisy, and unverified. They may include failed searches, repeated observations, abandoned hypotheses, weakly supported claims, and patch-construction details that should not be copied into the task specification. This creates a specification-refinement challenge: the method must recover useful repair specification evidence from the trajectory, keep high-level diagnoses connected to concrete repository observations, and review generated claims against the pre-fix repository before the refined report is given to a downstream repair agent.

We present \tool, a trajectory-guided approach for repository-supported specification refinement. Given an original bug report $b$ and a pre-fix repository snapshot $R_c$, \tool runs a trajectory-collection agent using only $b$ and $R_c$. As \tool does not validate any candidate patch produced during this trajectory-collection run, it discards any such patch and retains only the execution trajectory $\tau$. It then extracts specification evidence from $b$ and $\tau$, focusing on three repair-relevant dimensions: the failure mechanism, the behavioral requirement, and the implementation scope. 

To preserve both diagnostic structure and source-code evidence, \tool organizes this evidence into a hierarchical representation $M$ with progressively finer levels of detail: a high-level candidate interpretation, diagnostic relationships, and concrete repository observations. \tool then generates a draft refined report $\hat{b}$ from $b$ and $M$, and applies repository-based review against $R_c$ to remove unsupported claims, revise uncertain statements, and add repository-supported details that were omitted. The resulting report $\hat{b}_f$ serves as the repository-supported specification supplied to the downstream repair agent. This design separates trajectory use from patch trust. \tool does not assume that any candidate patch produced during the trajectory-collection run is correct, and it does not use the developer patch, post-fix repository, or benchmark outcome during specification refinement.

We evaluate \tool on all 300 SWE-Bench Lite~\cite{jimenez2024swebench} instances using Mini-SWE-Agent V2~\cite{minisweagentV2} as the primary downstream repair agent. With GPT-5-mini, \tool improves Pass@1 from 41.00\% using the original reports to 59.67\%. With MiniMax M2.5, it improves Pass@1 from 54.67\% to 64.33\%. To evaluate whether the refined reports remain useful beyond the primary downstream repair agent, we further evaluate \tool's refined reports on a stratified sample of 100 instances using Agentless~\cite{xia2025agentless} and AutoCodeRover~\cite{zhang2024autocoderover}. \tool improves Agentless from 41.00\% to 71.00\% and AutoCodeRover from 47.00\% to 72.00\%. Our ablation study shows that all the main components of \tool are important. Removing repository-based review reduces Pass@1 from 59.67\% to 48.00\%, while removing the hierarchical evidence representation reduces it to 47.67\%. These results show that trajectory-guided specification refinement with repository-based review provides more informative diagnostic context for repository-level repair.

In summary, this paper makes the following contributions:

\begin{itemize}
    \item We formulate bug report enhancement for repository-level APR as repository-supported specification refinement, where the goal is to make explicit the failure mechanism, behavioral requirement, and implementation scope needed for repair.

    \item We introduce \tool, a trajectory-guided approach that extracts and hierarchically organizes specification evidence from the trajectory produced by an unverified trajectory-collection run, reviews the evidence against the source code, and generates a refined report.

    \item We evaluate \tool on all 300 SWE-Bench Lite~\cite{jimenez2024swebench} instances using Mini-SWE-Agent V2~\cite{minisweagentV2} and show that its refined reports improve Pass@1 from 41.00\% to 59.67\% with GPT-5-mini and from 54.67\% to 64.33\% with MiniMax M2.5.

    \item We show that the benefits of \tool generalize across downstream repair agents. On a stratified sample of 100 instances, \tool improves Pass@1 from 41.00\% to 71.00\% for Agentless~\cite{xia2025agentless} and from 47.00\% to 72.00\% for AutoCodeRover~\cite{zhang2024autocoderover}.

    \item We conduct ablation and cost analysis showing that repository-based review and hierarchical evidence abstraction both substantially contribute to repair performance, while the additional preprocessing cost remains practical relative to the repair gains.
\end{itemize}

Our findings show a promising direction for improving repository-level repair: not only making repair agents stronger, but also improving the task specifications that guide them.

\phead{Paper Organization.} 
Section~\ref{sec:related} discusses related work. 
Section~\ref{sec:motivation} presents a motivating example. 
Section~\ref{sec:approach} describes the design of \tool. 
Section~\ref{sec:evaluation} reports the evaluation results. 
Section~\ref{sec:threats} discusses threats to validity. 
Finally, Section~\ref{sec:conclusion} concludes the paper.

\section{Related Work}\label{sec:related}

This section discusses prior work on bug-report enhancement, repository-level automated program repair (APR), and reuse of repair knowledge.

\phead{Bug Report Quality, Structuring, and Enhancement.}
Prior work has established that the information contained in a bug report affects how effectively developers can diagnose and resolve an issue. Developers particularly value reproduction steps, stack traces, and test cases~\cite{bettenburg2008makes}, and the presence of reproduction steps, stack traces, and fix suggestions can influence resolution time~\cite{soltani2020significance}. Building on these findings, subsequent work improves bug reports by reorganizing their existing content or recovering missing diagnostic information. Rastkar et al.~\cite{rastkar2010summarizing} summarize lengthy bug reports, while Acharya and Ginde~\cite{acharya2025can} transform unstructured reports into structured templates. LLPut~\cite{al2025llput} extracts failure-inducing inputs from report text, and Fahim et al.~\cite{fahim2025crash} enrich crash reports using stack traces and source-code context. These approaches improve the presentation or diagnostic content of a report using information available in the report and related artifacts. In contrast, \tool uses the trajectory produced by a trajectory-collection run to recover repair-relevant information that may not be explicit in the original report to better guide automated repair.

\phead{Enhancing Repository-Level APR.}
Recent approaches improve repository-level APR by clarifying the repair task, constructing repair guidance, or providing additional repository context for patch generation~\cite{suri2026codescout,zhang2026sgagent,tamoyan2026sherloc,yang2025enhancing,pan2026reporepair,kuang2026reagent}. CodeScout~\cite{suri2026codescout} enriches underspecified tasks through static repository pre-exploration before downstream repair. A broader line of work helps agents acquire and organize repair-relevant context without rewriting the issue. SWE-Agent~\cite{yang2024swe} and RepairAgent~\cite{bouzenia2025repairagent} support iterative repository exploration, tool use, patch generation, and validation. Agentless~\cite{xia2025agentless} separates localization, repair, and validation, whereas AutoCodeRover~\cite{zhang2024autocoderover} combines LLM reasoning with structure-aware code search and test-based localization. Other approaches produce explicit artifacts to guide repair: SGAgent~\cite{zhang2026sgagent} and SHERLOC~\cite{tamoyan2026sherloc} generate diagnostic guidance, while KGCompass~\cite{yang2025enhancing} and RepoRepair~\cite{pan2026reporepair} represent repository knowledge through knowledge graphs or hierarchical code documentation.

\tool differs in both its evidence source and its objective. CodeScout~\cite{suri2026codescout} derives task context from static repository pre-exploration, while agentic APR systems collect context as part of the final repair run. In contrast, \tool extracts specification evidence from the ordered trajectory of an unverified trajectory-collection run without assuming that any candidate patch from that run is correct. It hierarchically organizes this trajectory-derived evidence and reviews the resulting claims against the pre-fix repository before producing a refined report for downstream repair.

\phead{Reuse of Repair Knowledge.} 
Recent work studies how knowledge from previously resolved issues can guide repository-level repair. ExpeRepair~\cite{mu2025experepair} derives reusable memories and semantic insights from historical repair trajectories, while ConRAD~\cite{li2026outcome} reconstructs stage-wise repair reasoning from verified historical patches. SWE-ContextBench~\cite{zhu2026swe} evaluates whether coding agents can retrieve and reuse relevant context from related issues. These approaches transfer knowledge across issues. In contrast, \tool extracts specification evidence from the trajectory produced by an unverified trajectory-collection run for the current issue and uses it to refine the bug report before downstream repair.

\phead{\tool in Relation to Prior Work.} Overall, \tool connects bug-report enhancement with repository-level repair. Prior work typically improves reports using information already present in the report or related artifacts, or uses repository context directly during patch generation. \tool instead uses the trajectory produced by an unverified trajectory-collection run to uncover missing specification details and incorporates them into a refined report for downstream repair. It also differs from methods that reuse past repairs because the evidence comes from the current issue and does not depend on any patch from the trajectory-collection run being correct.

\section{A Motivating Example} 
\label{sec:motivation}

A bug report can describe a visible failure while omitting repair-relevant behavior that is evident in the repository. Consider \texttt{astropy-14365}, a real-world GitHub issue from the Astropy project included in the SWE-Bench Lite benchmark~\cite{jimenez2024swebench}. Figure~\ref{fig:qdp-report} presents simplified excerpts from the original bug report and the refined report produced by \tool. The original report provides two useful pieces of specification information: the observed failure, in which the reader rejects a lowercase QDP command, and the expected behavior, in which commands should be accepted regardless of capitalization. However, the original report does not explain the specific code behavior responsible for the failure or identify the full scope of the parsing logic affected by the same case-sensitivity assumption.

\begin{figure*}[t]
\centering
\scriptsize

\newlength{\qdpfigheight}
\setlength{\qdpfigheight}{5.45cm}

\begin{minipage}[t][\qdpfigheight][t]{0.31\textwidth}
\hrule
\vspace{0.5em}

\textbf{Original bug report excerpt}

\medskip
\textbf{Reported issue:} QDP commands are incorrectly case-sensitive.

\medskip
\textbf{Expected behavior.}
Because QDP commands are case-insensitive, the \texttt{ascii.qdp} reader should accept both
\texttt{READ SERR 1 2} and \texttt{read serr 1 2}.

\medskip
\textbf{Observed behavior.}
Reading a file containing the lowercase command

\begin{lstlisting}[
language={},
basicstyle=\ttfamily\scriptsize,
columns=fullflexible,
aboveskip=0.2em,
belowskip=0.2em
]
read serr 1 2
1 0.5 1 0.5
\end{lstlisting}

causes the reader to fail with

\begin{lstlisting}[
language={},
basicstyle=\ttfamily\scriptsize,
breaklines=true,
columns=fullflexible,
aboveskip=0.2em,
belowskip=0.2em
]
ValueError: Unrecognized
QDP line: read serr 1 2
\end{lstlisting}

\vfill
\hrule
\end{minipage}
\hfill
\begin{minipage}[t][\qdpfigheight][t]{0.65\textwidth}
\hrule
\vspace{0.5em}

\textbf{\tool refined report excerpt}

\medskip
\textbf{Title.}
\texttt{ascii.qdp} reader treats QDP commands and masked token \texttt{NO} as case-sensitive.

\medskip
\textbf{Description.}
The reader recognizes QDP commands and masked values in a case-sensitive way. Lowercase or mixed-case commands such as \texttt{read serr 1 2}, and masked tokens such as \texttt{no}, may therefore cause parsing failures or incorrect data interpretation.

\medskip
\textbf{RootCause.}
The issue is caused by uppercase-only assumptions in related parsing logic:
(1) command matching in \texttt{\_line\_type} is compiled without case-insensitive matching, and
(2) masked-value parsing in \texttt{\_get\_tables\_from\_qdp\_file} checks only the exact token \texttt{NO}.

\medskip
\textbf{StepsToReproduce.}
Create \texttt{test.qdp} containing \texttt{read serr 1 2} and \texttt{1 0.5 1 0.5}, then run \texttt{Table.read('test.qdp', format='ascii.qdp')}.

\medskip
\textbf{ExpectedBehavior.}
QDP control commands such as \texttt{read serr 1 2} and masked-value tokens such as \texttt{no}, \texttt{No}, or \texttt{NO} should be accepted irrespective of case.

\medskip
\textbf{ObservedBehavior.}
The reader raises \texttt{ValueError} for lowercase commands, and lowercase or mixed-case masked tokens are not treated as masked values.

\vfill
\hrule
\end{minipage}

\caption{Simplified before-and-after report excerpts for \texttt{astropy-14365}. The original report describes the visible lowercase-command failure, while the refined report preserves the reported failure and expected behavior and adds repository-supported information about the failure mechanism, behavioral requirement, and implementation scope.}
\label{fig:qdp-report}
\end{figure*}

The developer repair, shown in simplified form in Listing~\ref{lst:qdp-repair}, illustrates why the additional information in the refined report is useful. The repair modifies two locations. First, it makes command classification in \texttt{\_line\_type} case-insensitive, directly addressing the reported failure. Second, it makes the handling of the masked-value token \texttt{NO} case-insensitive in \texttt{\_get\_tables\_from\_qdp\_file}. This second location is not mentioned in the original report, although it embodies the same uppercase-only parsing assumption.

\begin{lstlisting}[
language=Python,
caption={Simplified developer repair for \texttt{astropy-14365}. The first change addresses the reported command failure, while the second applies the same case-insensitive behavior to a related parsing location omitted in the original report.},
label={lst:qdp-repair},
basicstyle=\ttfamily\footnotesize,
breaklines=true,
columns=fullflexible
]
# Reported behavior: command classification in _line_type
# Before: matching is case-sensitive
_line_type_re = re.compile(_type_re)
# After: commands are matched regardless of case
_line_type_re = re.compile(_type_re, re.IGNORECASE)

# Related behavior: masked-value parsing in _get_tables_from_qdp_file
# Before: only uppercase "NO" is recognized
if v == "NO":
    values.append(np.ma.masked)
# After: the token is recognized regardless of case
if v.upper() == "NO":
    values.append(np.ma.masked)
\end{lstlisting}

A repair agent guided only by the original report is naturally directed toward the visible command-classification failure. It may correctly modify the regular expression so that \texttt{read serr 1 2} matches the same parser rule as \texttt{READ SERR 1 2}. Such a repair addresses the reported symptom but can remain incomplete: the report does not indicate that the comparison \texttt{v == "NO"} represents a related instance of the same case-sensitive parsing behavior. Discovering this second location, therefore, requires reasoning beyond the explicit contents of the report. Indeed, the developer patch modifies both locations, showing that the intended repair extends beyond the failure explicitly described in the report.

These missing specification elements could guide the repair agent toward a more complete fix, but they are not available in the original report. A trajectory-collection run can expose such elements because its trajectory records the agent's code inspections, search decisions, and intermediate reasoning while it tries to explain the reported failure. In this example, the trajectory shows how reasoning starts from the reported command-matching failure and moves to related parsing logic that contains the \texttt{v == "NO"} comparison. \tool therefore analyzes the trajectory produced by an unverified trajectory-collection run to recover evidence about the failure mechanism, broader behavioral requirement, and affected code locations. It then structures and reviews this evidence against the pre-fix repository to produce a refined report for downstream repair.

\section{\tool: Trajectory-Guided Specification Refinement}
\label{sec:approach}

\begin{figure*}
\centering
\includegraphics[width=0.9\textwidth]{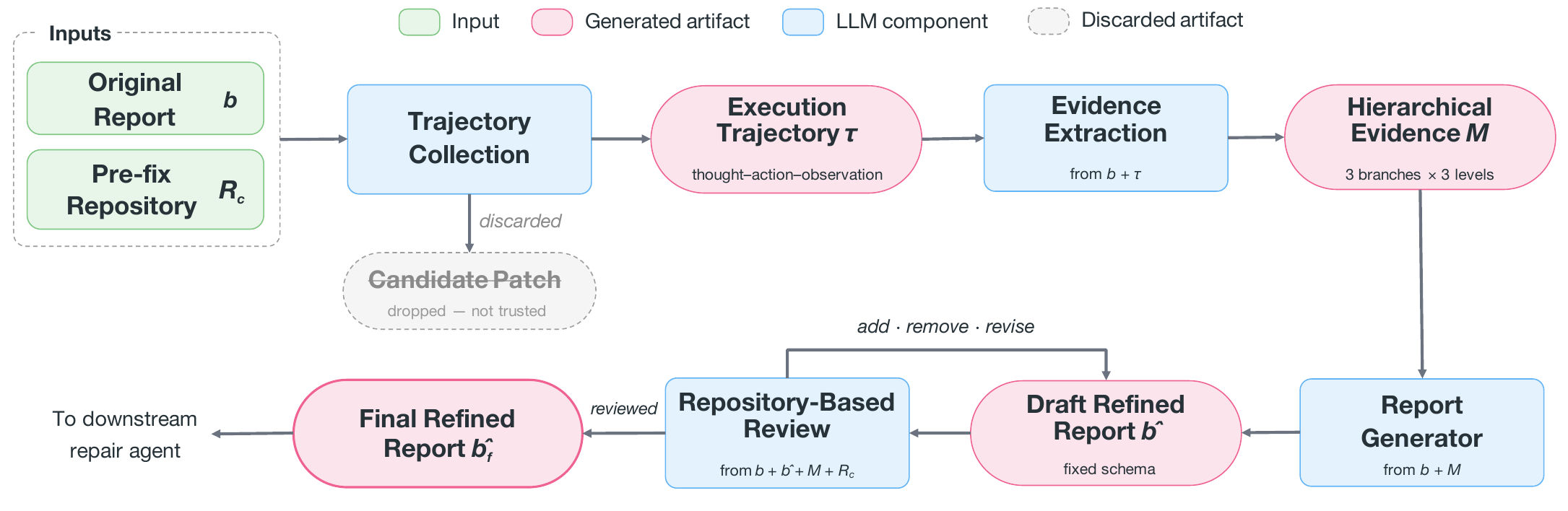}
\caption{\textbf{Overview of \tool.} Given an original bug report $b$ and the pre-fix repository snapshot $R_c$, \tool produces a final refined report $\hat{b}_f$ that serves as a repository-supported specification for downstream repair. \tool first runs a trajectory-collection agent using only $b$ and $R_c$, discards any candidate patch produced during this run, and retains the execution trajectory $\tau$. It then uses $b$ and $\tau$ to recover specification evidence and structure it into a hierarchical evidence representation $M$. Finally, \tool generates a draft refined report $\hat{b}$ from $b$ and $M$, and applies repository-based review using $b$, $\hat{b}$, $M$, and $R_c$ to produce the final refined report $\hat{b}_f$.}
\label{fig:framework}
\end{figure*}

\tool addresses the problem of underspecified bug reports by using the trajectory produced by an unverified trajectory-collection run to recover trajectory-derived specification evidence. Given an original bug report $b$ and its pre-fix repository snapshot $R_c$, \tool runs a trajectory-collection agent using only $b$ and $R_c$, and retains the resulting execution trajectory $\tau$. The trajectory records the agent's repository searches, inspected source code, tool invocations, observations, and intermediate reasoning. From $b$ and $\tau$, \tool extracts repair-relevant observations about three missing specification elements: the failure mechanism, the behavioral requirement, and the affected implementation scope. It organizes these observations into a hierarchical specification $M$, generates a draft refined report $\hat{b}$, and reviews the draft against the pre-fix repository $R_c$ to remove unsupported claims, revise uncertain statements, and add repository-supported details. The resulting report $\hat{b}_f$ is supplied to the downstream repair agent as the task specification for final patch generation. Figure~\ref{fig:framework} provides an overview of \tool.

\subsection{Collecting Specification Evidence from a Trajectory-Collection Run}
\label{exploration}
To recover specification missing from the original report, \tool performs an unverified trajectory-collection run on the original bug report $b$ and pre-fix repository snapshot $R_c$ using a trajectory-collection agent. The agent performs repository-level exploration for the reported issue: it searches the repository, inspects source code, executes available tools, reasons about candidate failure explanations, and records each thought-action-observation step. This run produces the execution trajectory $\tau$ and may also produce a candidate patch. \tool, however, is agnostic to the specific trajectory source: it requires only $\tau$, not the candidate patch or the agent's internal implementation, so any repository-level agent that exposes such a trajectory can potentially serve as the trajectory source. To avoid data leakage, the trajectory-collection agent has no access to the developer patch, post-fix code, or benchmark test outcome. We call the trajectory-collection run \emph{unverified} because \tool does not validate whether any resulting patch is correct during specification refinement. Any candidate patch is therefore discarded, and only $\tau$ is retained. We keep $\tau$ because it records the agent's repository interactions, intermediate reasoning, and repair decisions, which can provide trajectory-derived specification evidence regardless of whether the patch was correct.

We represent $\tau$ as an ordered sequence of thought--action--observation tuples~\cite{agent_traj_thought_action_result}:
\[
\tau=\langle (t_1,a_1,o_1), \ldots, (t_n,a_n,o_n)\rangle,
\]
where each tuple records one agent-environment interaction step. The thought $t_i$ records the agent's stated diagnostic reasoning, $a_i$ denotes the corresponding repository or tool action, and $o_i$ denotes the resulting observation. An action may search for a program entity, inspect or modify source code, or execute an available tool. An observation may contain search results, retrieved source code, execution feedback, or other tool output. The observation $o_i$ can inform subsequent reasoning and actions, allowing the ordered trajectory to preserve how the agent's interpretation and candidate repair evolve during the run.

Although the run is unverified, its trajectory can still expose information useful for refining the report. It records how the trajectory-collection agent attempts to resolve ambiguities in the original report and construct a working specification of the issue. This information may include the failure behavior observed by the agent, the files, methods, or APIs it associates with the failure, the inputs or configurations under which the behavior occurs, behavior suggested by analogous implementations, and constraints that a final repair should preserve. At the same time, the trajectory may contain irrelevant searches, unsupported hypotheses, abandoned diagnoses, weakly supported conclusions, and patch-construction details that do not contribute to the specification. \tool therefore treats $\tau$ as a noisy source of trajectory-derived specification evidence, rather than as a specification or trusted repair, and passes it to the hierarchical evidence-abstraction step.

\subsection{Hierarchical Specification Evidence Abstraction}
\label{evidence-abstraction}

The execution trajectory $\tau$ produced in \S\ref{exploration} can be lengthy and noisy. Passing it directly to specification generation may obscure relevant findings, whereas compressing it into a flat summary may discard repository details needed to support the generated specification. \tool therefore transforms the original bug report $b$ and trajectory $\tau$ into a hierarchical evidence representation $M$. It first extracts candidate repository findings $E$ and then organizes them into evidence structures supporting three aspects of the missing specification: the failure mechanism, behavioral requirement, and implementation scope.

\begin{table*}[t]
\centering
\small
\caption{An illustrative example of the hierarchical evidence representation.}
\label{tab:qdp-hierarchical-evidence}
\begin{tabular}{p{0.07\textwidth}p{0.28\textwidth}p{0.28\textwidth}p{0.28\textwidth}}
\toprule
\textbf{Level} &
\textbf{Failure mechanism} &
\textbf{Behavioral requirement} &
\textbf{Implementation scope} \\
\midrule

High &Case-sensitive QDP parsing &Accept tokens regardless of capitalization &Apply the behavior consistently across the reader \\

Mid &Multiple parser checks assume uppercase input &Handle commands and masked values case-insensitively &Update both command and masked-token parsing \\

Low &Missing regex flag and exact \texttt{"NO"} comparison &Recognize lowercase commands and masked tokens &Relevant functions and tests in \texttt{qdp.py} \\

\bottomrule
\end{tabular}
\end{table*}

\phead{Specification Evidence Extraction.}
\tool uses a structured LLM-based abstraction prompt to extract candidate findings from $b$ and $\tau$. Rather than summarizing the trajectory chronologically, the prompt retains observations that help reconstruct: \emph{1) the failure mechanism}, or the source-code behavior that may explain the reported symptom; \emph{2) the behavioral requirement}, or the behavior that should hold in the reported scenario; and \emph{3) the implementation scope}, or the code locations that implement, depend on, or otherwise participate in that behavior. We denote the extracted findings and repository observations as $E$.

The LLM omits routine tool interactions, repeated observations, failed lookups, and details unrelated to understanding the issue. It retains intermediate findings when they support a candidate specification claim, connect multiple code locations, or reveal an unresolved gap. Because $\tau$ is produced by an unverified trajectory-collection run, however, $E$ may contain weakly supported claims, incomplete evidence, or conflicting interpretations. \tool therefore treats the extracted findings as candidates to be examined during the subsequent repository-based review.

\phead{Hierarchical Evidence Representation.}
\tool organizes $E$ into three evidence branches:
\[
M=
\left(
M_{\mathrm{mech}},
M_{\mathrm{req}},
M_{\mathrm{scope}}
\right),
\]
where $M_{\mathrm{mech}}$, $M_{\mathrm{req}}$, and $M_{\mathrm{scope}}$ contain evidence concerning the candidate failure mechanism, behavioral requirement, and implementation scope, respectively. Each branch is represented at three levels:
\[
M_j=
\left(
M_{j,\mathrm{hi}},
M_{j,\mathrm{mid}},
M_{j,\mathrm{lo}}
\right),
\qquad
j\in{\mathrm{mech},\mathrm{req},\mathrm{scope}}.
\]

Table~\ref{tab:qdp-hierarchical-evidence} gives an illustrative example of this representation using the QDP motivating example based on Figure~\ref{fig:qdp-report}. In this representation, for each specification aspect $j$, $M_{j,\mathrm{hi}}$ states the candidate specification-level conclusion. $M_{j,\mathrm{mid}}$ captures the diagnostic reasoning that supports the conclusion, such as relationships among code paths, shared behavioral patterns, or dependencies between affected locations. $M_{j,\mathrm{lo}}$ retains the concrete repository observations that support this reasoning, including relevant files, functions, variables, conditions, and constants. Thus, the three branches distinguish the content of the reconstructed specification, while the three levels preserve how repository observations support specification-level conclusions.

\subsection {Specification Generation and Repository-Based Review}
\label{sec:specification-generation}
The hierarchical evidence representation, $M$, organizes candidate specification evidence, but it is not yet a refined report. Its information remains distributed across high-level interpretations, diagnostic relationships, and concrete repository observations. Because this evidence comes from a useful but unverified trajectory, \tool must separate candidate report content from unsupported assumptions before giving the refined report to a downstream repair agent. \tool therefore converts $M$, together with the original bug report $b$, into a draft refined report $\hat{b}$, and then applies repository-based review against the pre-fix repository $R_c$ to produce the final refined report $\hat{b}_f$.

\phead{Draft Specification Generation.}
Given the original bug report $b$ and hierarchical evidence representation $M$, \tool invokes an LLM to generate a draft refined report $\hat{b}$. The report follows a fixed schema with fields that prior work has found useful for bug understanding, diagnosis, and reproduction~\cite{bettenburg2008makes,soltani2020significance,hirsch2020root,medeiros2024impact,fahim2025crash}. The schema includes \emph{Title}, \emph{Description}, \emph{RootCause}, \emph{StepsToReproduce}, \emph{ExpectedBehavior}, and \emph{ObservedBehavior}.

The fields play distinct roles. \emph{Title} and \emph{Description} summarize the reported issue and its manifestation. \emph{ObservedBehavior} describes the behavior exhibited by the pre-fix implementation. \emph{ExpectedBehavior} states the behavioral requirement that should hold for the reported scenario. \emph{RootCause} records the candidate failure mechanism, implementation scope, and repository-supported details that explain the observed failure. \emph{StepsToReproduce} is populated only when the original report $b$ or hierarchical evidence representation $M$ supports concrete reproduction steps. This schema separates the externally visible failure from the repository-level diagnosis and keeps the report focused on the issue rather than on a prescribed code change.

\phead{Specification Review and Report Revision.}
Because both the draft report $\hat{b}$ and its supporting evidence representation $M$ ultimately depend on the unverified trajectory $\tau$, \tool performs a repository-based review on $\hat{b}$ before producing the final refined report $\hat{b}_f$. The reviewer is an LLM-based component provided with the original report $b$, the draft report $\hat{b}$, and the hierarchical evidence representation $M$. It also has read-only tools for searching and inspecting the pre-fix repository $R_c$.

The reviewer uses $M$ as a structured index of candidate claims, diagnostic relationships, and relevant source locations. For each substantive claim introduced in $\hat{b}$, the reviewer assesses whether the claim is supported by the evidence recorded in $M$ and, where necessary, inspects the referenced source code in $R_c$ to validate that support. The review considers four properties: whether added claims have adequate repository support, whether relevant findings in $M$ have been omitted, whether the stated implementation scope is appropriately bounded, and whether each claim is expressed with certainty proportional to the available evidence.

The reviewer gives particular attention to behavioral requirements. Requirements explicitly stated in the original report $b$ are preserved as reported expectations, whereas requirements reconstructed from $M$ are treated as inferred claims. An inferred requirement is retained only when it is supported by repository behavior observed in $R_c$, and is expressed as the behavior that should hold rather than as verified post-fix behavior. 

When evidence is insufficient, the reviewer removes the claim, revises its wording to reflect uncertainty, or limits it to the supported scope. When the draft report $\hat{b}$ omits a relevant repository-supported finding, the reviewer adds it. Unresolved gaps remain marked as uncertain rather than being converted into definitive claims. This review improves the evidential support, coverage, and internal consistency of $\hat{b}$, but it does not establish ground-truth correctness because neither the developer patch nor the benchmark test outcome is available.

The resulting report $\hat{b}_f$ is then provided to the downstream repair agent for repair.

\section{Evaluation}
\label{sec:evaluation}

We evaluate \tool on SWE-Bench Lite~\cite{jimenez2024swebench}, a 300-instance subset of SWE-Bench that provides a more cost-efficient evaluation of repository-level APR. Each instance is derived from a real-world GitHub issue and includes the corresponding issue report, a pre-fix repository snapshot, and a test oracle for validating generated patches. We measure repair performance using Pass@1, the percentage of instances for which the first generated patch passes the test oracle. Our four research questions evaluate downstream repair effectiveness, cross-agent generalization, component contributions, and computational cost.

\subsection*{\RQOne}
\label{RQ1}

\begin{table*}
\centering
\caption{Repair performance of Mini-SWE-Agent V2 using original reports, \baseline reports, and \tool's refined reports. Results are reported as resolved bugs and Pass@1 percentages. For each repository and LLM backbone, the best-performing report variant is highlighted in \textbf{bold}.}
\label{tab:rq1_minisweagent}
{
\setlength{\tabcolsep}{12pt}
\scalebox{1.00}{
\begin{tabular}{l|rrr|rrr}
\toprule
\multirow{2}{*}{\textbf{Repository}} 
& \multicolumn{3}{c|}{\textbf{GPT-5-mini}} 
& \multicolumn{3}{c}{\textbf{MiniMax M2.5}} \\
\cmidrule(lr){2-4} \cmidrule(lr){5-7}
& \textbf{Original} 
& \textbf{\baseline} 
& \textbf{\tool} 
& \textbf{Original} 
& \textbf{\baseline} 
& \textbf{\tool} \\
\midrule
astropy (6)
& 3 (50.00\%)
& 4 (66.67\%)
& \textbf{5 (83.33\%)}
& 3 (50.00\%)
& 4 (66.67\%)
& \textbf{5 (83.33\%)} \\

django (114)
& 58 (50.88\%)
& 69 (60.53\%)
& \textbf{82 (71.93\%)}
& 72 (63.16\%)
& 71 (62.28\%)
& \textbf{83 (72.80\%)} \\

matplotlib (23)
& 10 (43.48\%)
& 11 (47.83\%)
& \textbf{14 (60.87\%)}
& 11 (47.83\%)
& 11 (47.83\%)
& \textbf{13 (56.52\%)} \\

seaborn (4)
& 1 (25.00\%)
& 1 (25.00\%)
& \textbf{2 (50.00\%)}
& \textbf{3 (75.00\%)}
& 2 (50.00\%)
& \textbf{3 (75.00\%)} \\

flask (3)
& 0 (0.00\%)
& 0 (0.00\%)
& 0 (0.00\%)
& 0 (0.00\%)
& 0 (0.00\%)
& \textbf{1 (33.33\%)} \\

requests (6)
& 2 (33.33\%)
& 3 (50.00\%)
& \textbf{6 (100.00\%)}
& 3 (50.00\%)
& 3 (50.00\%)
& \textbf{6 (100.00\%)} \\

xarray (5)
& 1 (20.00\%)
& 1 (20.00\%)
& \textbf{2 (40.00\%)}
& \textbf{2 (40.00\%)}
& \textbf{2 (40.00\%)}
& \textbf{2 (40.00\%)} \\

pylint (6)
& 1 (16.67\%)
& 1 (16.67\%)
& \textbf{2 (33.33\%)}
& \textbf{2 (33.33\%)}
& \textbf{2 (33.33\%)}
& \textbf{2 (33.33\%)} \\

pytest (17)
& 3 (17.65\%)
& \textbf{4 (23.53\%)}
& \textbf{4 (23.53\%)}
& 7 (41.18\%)
& 7 (41.18\%)
& \textbf{9 (52.94\%)} \\

scikit-learn (23)
& 11 (47.83\%)
& 12 (52.17\%)
& \textbf{14 (60.87\%)}
& 14 (60.87\%)
& 13 (56.52\%)
& \textbf{17 (73.91\%)} \\

sphinx (16)
& 8 (50.00\%)
& 8 (50.00\%)
& \textbf{10 (62.50\%)}
& \textbf{9 (56.25\%)}
& 8 (50.00\%)
& \textbf{9 (56.25\%)} \\

sympy (77)
& 25 (32.47\%)
& 32 (41.56\%)
& \textbf{38 (49.35\%)}
& 38 (49.35\%)
& 36 (46.75\%)
& \textbf{43 (55.84\%)} \\

\midrule
\textbf{Total (300)}
& 123 (41.00\%)
& 146 (48.67\%)
& \textbf{179 (59.67\%)}
& 164 (54.67\%)
& 159 (53.00\%)
& \textbf{193 (64.33\%)} \\
\bottomrule
\end{tabular}}}
\end{table*}

\phead{Motivation.}
Repository-supported specification refinement is useful only if the resulting refined bug report provides actionable guidance for resolving the underlying bug. In our setting, this guidance connects the observed failure to its failure mechanism, the behavioral requirement that should hold, and the implementation scope involved. Because repair agents use the bug report as the initial problem description for generating a patch, automated program repair provides a natural downstream evaluation of whether \tool's refined bug reports contain diagnostic information that helps generate correct patches. RQ1 therefore evaluates the extent to which \tool improves repair performance through repository-supported specification refinement.

\phead{Approach.} For each bug instance, we evaluate the repair performance obtained using \tool's refined reports against two baseline report variants: the original reports and the \baseline reports. Note that \textit{the downstream repair agent remains completely identical}, and the only difference is the bug report we use as input. 

\noindent{\textit{\textbf{Original report:}}} The bug report provided by SWE-Bench Lite~\cite{jimenez2024swebench}.

\noindent{\textit{\textbf{\baseline report}}:} A trajectory-based report-enhancement baseline. We use this trajectory-based baseline, rather than a report-only rewrite baseline, because prior work shows that additional task evidence improves report enhancement and repair guidance~\cite{zhu2026specification, huang2025dynafix, fahim2025crash, akyol2026improbr}. Given the original bug report $b$, this baseline uses the same trajectory-collection agent as \tool to collect the raw execution trajectory $\tau$. The report-generation model then receives $b$ and $\tau$ directly to produce the raw refined report $\hat{b}_{\mathrm{raw}}$. Unlike \tool, this baseline does not construct the hierarchical evidence representation $M$ or perform repository-based review.

\noindent{\textit{\textbf{\tool report:}}} The refined bug report generated by \tool following the repository-supported specification refinement approach described in Section~\ref{sec:approach}, including hierarchical evidence abstraction and repository-based review.

We select \emph{Mini-SWE-Agent V2}~\cite{minisweagentV2} as our primary downstream repair agent because it provides a lightweight yet capable workflow for repository-level automated program repair and has been commonly used in prior studies~\cite{pei2025scope, lyu2026agentszz, yang2026twinrouterbench, li2026dualeval, joshi2026inference, zhang2026coda}. To examine whether the effect of report refinement is consistent across model backbones, we run Mini-SWE-Agent V2 with two LLMs: GPT-5-mini (\texttt{gpt-5-mini-2025-08-07})~\cite{gpt-5-mini} and MiniMax M2.5 (\texttt{minimax-m2.5})~\cite{minimax-m2.5}. For each LLM setting, the generated reports are produced with the same backbone used for downstream repair: GPT-5-mini performs all LLM calls in the GPT-5-mini setting, including trajectory collection, evidence extraction, hierarchical evidence abstraction, draft report generation, and repository-based review, while MiniMax M2.5 performs the same components in the MiniMax M2.5 setting. We select these two LLMs because they provide strong coding and agentic capabilities at practical API cost: GPT-5-mini is designed for cost-sensitive, high-volume workloads, while MiniMax M2.5 targets coding and agentic tasks with low token pricing~\cite{gpt-5-mini,minimax-m2.5}.

\phead{Results.} 
\noindent\textbf{\em \tool reports improve Pass@1 over the original reports by 45.53\% with GPT-5-mini and 17.68\% with MiniMax M2.5.}
Table~\ref{tab:rq1_minisweagent} reports the Pass@1 repair results. Since the downstream repair agent remains unchanged, the differences in Pass@1 reflect the effect of changing the report given to the agent. With GPT-5-mini, Mini-SWE-Agent V2 achieves 41.00\% Pass@1 using the original reports and 48.67\% using \baseline reports. \tool reports increase Pass@1 to 59.67\%, corresponding to a 45.53\% relative improvement over the original reports and a 22.60\% relative improvement over \baseline reports. With MiniMax M2.5, the original reports already provide a stronger starting point, achieving 54.67\% Pass@1, while \baseline reports achieve 53.00\%. \tool reports further increase Pass@1 to 64.33\%, corresponding to a 17.68\% relative improvement over the original reports and a 21.38\% relative improvement over \baseline reports. This contrast shows that \baseline reports improve Pass@1 in one setting but not the other, whereas \tool reports improve repair performance with both LLM backbones.

\noindent\textbf{\em \tool reports consistently improve repair performance across repositories.}
The repository-level results show that \tool reports improve repair performance across repositories. With GPT-5-mini, \tool reports outperform the original reports in 11 of 12 repositories and \baseline reports in 10 repositories. Flask instances are unresolved by all report variants, while \tool reports tie \baseline reports on pytest. A similar pattern appears with MiniMax M2.5. \tool reports achieve the best or tied-best Pass@1 in every repository, improving over the original reports in eight repositories and over \baseline reports in 10 repositories. These results show that the benefit of \tool's repository-supported refined reports is broadly distributed across projects and is not driven by gains in only a small number of repositories.

\begin{figure}[t]
\centering
\includegraphics[width=\columnwidth]{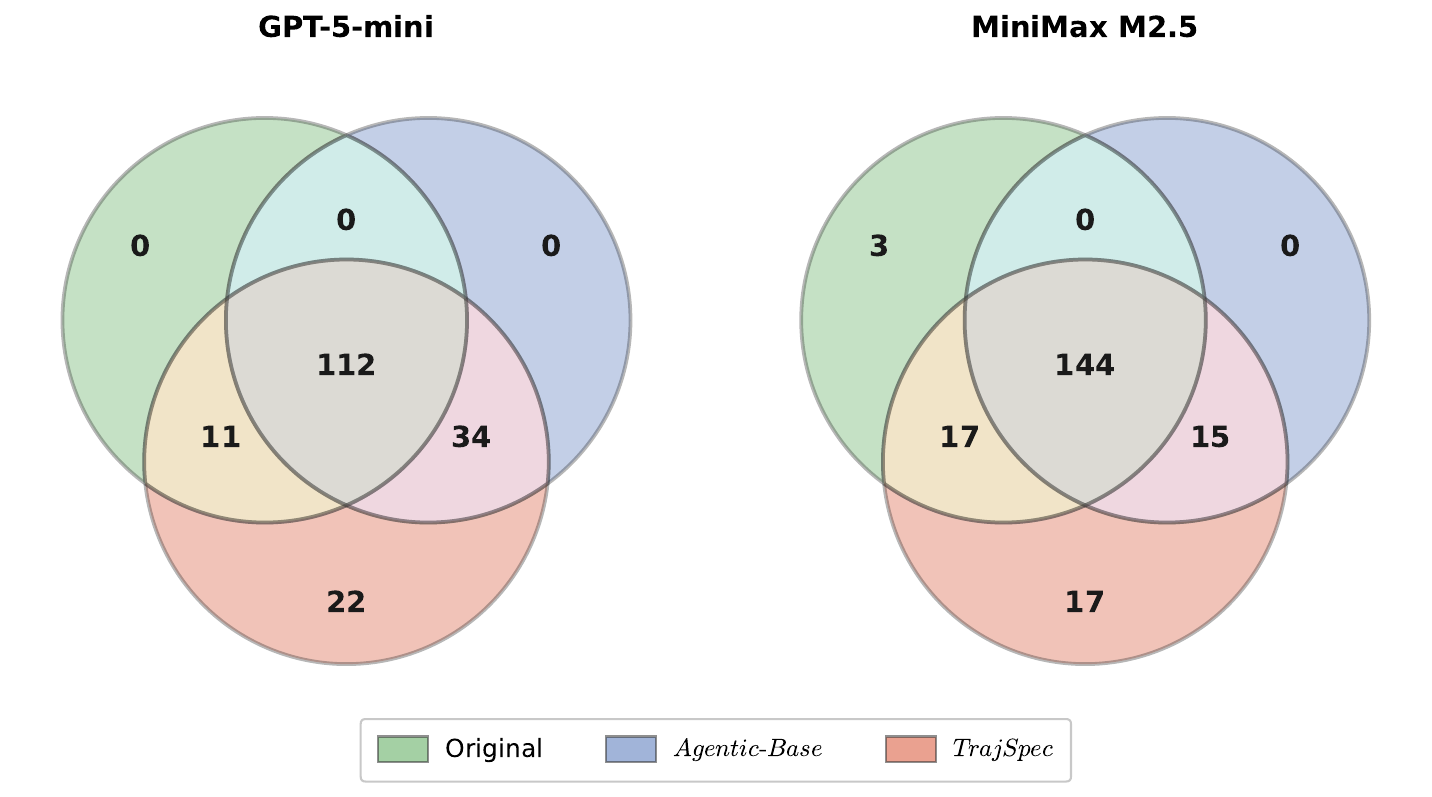}
\caption{Overlap of repaired SWE-Bench Lite instances across the three report variants.}
\label{fig:overlap_venn}
\end{figure}

\noindent\textbf{\em \tool reports expand repair coverage while preserving almost all previously repaired instances.}
Figure~\ref{fig:overlap_venn} compares the instances repaired using the three report variants. With GPT-5-mini, every instance repaired using either the original reports or \baseline reports is also repaired using \tool reports. In total, \tool reports repair 179 instances ($11+112+34+22$), including 22 instances that neither comparison report repairs. Thus, with GPT-5-mini, the gains from \tool reports come entirely from expanding repair coverage rather than replacing previously repaired instances.

A similar pattern appears with MiniMax M2.5. In total, \tool reports repair 193 instances ($17+144+15+17$). They preserve 161 of the 164 instances repaired using the original reports ($17+144$ out of $3+17+144$) and all 159 instances repaired using \baseline reports ($144+15$). At the same time, \tool reports repair 32 additional instances beyond the original reports ($15+17$), including 17 instances that neither comparison report repairs. The three regressions are resolved upon rerunning, suggesting that they result from run-to-run variation rather than a systematic loss caused by the refined reports. These results show that \tool reports mostly expand the set of repaired bugs while preserving almost all instances repaired by the original or \baseline reports.

\rqboxc{
\tool reports improve Pass@1 over the original reports by 45.53\% with GPT-5-mini and 17.68\% with MiniMax M2.5. They also outperform \baseline reports under both LLM backbones and expand repair coverage while preserving almost all previously repaired instances.}

\subsection*{\RQTwo}
\label{RQ2}

\phead{Motivation.}
RQ1 evaluates \tool reports using Mini-SWE-Agent V2~\cite{minisweagentV2} as the downstream repair agent. The observed gains may therefore depend on how Mini-SWE-Agent V2 uses the refined reports, rather than on whether the same reports are useful for other repair agents. RQ2 addresses this question by providing the same reports to other repository-level repair agents and measuring their repair performance.

\phead{Approach.}
We reuse the GPT-5-mini-generated report variants evaluated in RQ1, including the original reports, \baseline reports, and \tool reports, and provide them to Agentless~\cite{xia2025agentless} and AutoCodeRover~\cite{zhang2024autocoderover}. These two repair agents differ from Mini-SWE-Agent V2 in how they organize repository navigation, localization, and patch generation. Running both agents on all 300 SWE-Bench Lite~\cite{jimenez2024swebench} instances for each report variant would require substantially greater computational resources. We therefore evaluate them on a stratified sample of 100 instances from SWE-Bench Lite, preserving the proportion of instances from each repository in the full 300-instance dataset. For each report variant, we run Agentless and AutoCodeRover independently using GPT-5-mini (\texttt{gpt-5-mini-2025-08-07})~\cite{gpt-5-mini} as the underlying language model and report the Pass@1.

\phead{Results.} 
\noindent\textbf{\em \tool reports improve Pass@1 over the original reports by 73.17\% with Agentless and 53.19\% with AutoCodeRover.}
Table~\ref{tab:rq2_generalization} reports the Pass@1 results for Agentless and AutoCodeRover. With Agentless, the original reports achieve 41.00\% Pass@1, while \baseline reports increase Pass@1 to 54.00\%. \tool reports further increase Pass@1 to 71.00\%, corresponding to a 73.17\% relative improvement over the original reports and a 31.48\% relative improvement over \baseline reports. With AutoCodeRover, the original reports achieve 47.00\% Pass@1 and \baseline reports achieve 52.00\%. \tool reports an increase in Pass@1 to 72.00\%, corresponding to a 53.19\% relative improvement over the original reports and a 38.46\% relative improvement over \baseline reports.

These improvements appear across two repair agents with different designs. Agentless decomposes repair into localization, patch generation, and validation, while AutoCodeRover uses structure-aware repository search to guide patch generation. The consistent gains across these agents suggest that the benefit of trajectory-guided specification refinement is not tied to the workflow of a particular downstream repair agent, but extends to agents with different localization and patch-generation strategies.

\begin{table}[t]
\centering
\caption{Pass@1 of Agentless and AutoCodeRover using original reports, \baseline reports, and \tool reports. The best result for each repair agent is highlighted in \textbf{bold}.}
\label{tab:rq2_generalization}
{
\setlength{\tabcolsep}{16pt}
\begin{tabular}{lcc}
\toprule
\textbf{Report Variant} & \textbf{Agentless} & \textbf{AutoCodeRover} \\ 
\midrule
Original      & 41\% & 47\% \\
Agentic-Base   & 54\% & 52\% \\
\tool         & \textbf{71\%} & \textbf{72\%} \\
\bottomrule
\end{tabular}}
\end{table}

\rqboxc{
On 100 stratified SWE-Bench Lite instances, \tool reports improve Pass@1 over the original reports by 73.17\% with Agentless and 53.19\% with AutoCodeRover. They also outperform \baseline reports under both downstream repair agents.}

\subsection*{\RQThree}
\label{RQ3}

\phead{Motivation.}
\tool relies on two key components to produce repository-supported specifications: a hierarchical evidence representation and repository-based review. In this RQ, we ablate the hierarchical evidence representation and repository-based review to quantify their contributions to \tool's repair performance.

\phead{Approach.} 
We construct two ablated variants of \tool, each removing one key component:

\noindent{\textit{\textbf{1) Without repository-based review:}}} This variant omits the repository-based reviewer agent. The draft refined report is used directly as input to the downstream repair agent, without repository-based review or revision.

\noindent{\textit{\textbf{2) Without hierarchical evidence representation:}}} This variant removes the hierarchical evidence representation. Therefore, it generates the refined report without organizing the extracted evidence into the high-level, mid-level, and low-level structure described in Section~\ref{evidence-abstraction}, while keeping all other components unchanged in \tool.

We conduct our ablation study on all 300 SWE-Bench Lite~\cite{jimenez2024swebench} instances. For \tool and both ablated variants, we use GPT-5-mini (\texttt{gpt-5-mini-2025-08-07})~\cite{gpt-5-mini} for every LLM call involved in the corresponding refinement variant, including trajectory collection, evidence extraction, hierarchical evidence abstraction when present, draft report generation, and repository-based review when present. We then use Mini-SWE-Agent V2~\cite{minisweagentV2} with GPT-5-mini to generate patches from the resulting reports and measure Pass@1 repair success.

\begin{table}[t]
\centering
\caption{Ablation results on 300 SWE-Bench Lite instances using Mini-SWE-Agent V2 with GPT-5-mini. Results are reported as the number of resolved instances and Pass@1. Arrows indicate relative decrease compared with the full \tool.}
\label{tab:rq3_ablation_summary}
{
\setlength{\tabcolsep}{18pt}
\scalebox{1.0}{
\begin{tabular}{lrr}
\toprule
\textbf{Variant} & \textbf{Resolved} & \textbf{Pass@1} \\
\midrule
\tool & \textbf{179} & \textbf{59.67\%} \\
w/o Reviewer & 144 & 48.00\% ($\downarrow$19.55\%) \\
w/o Hierarchy & 143 & 47.67\% ($\downarrow$20.11\%) \\
\bottomrule
\end{tabular}}}
\end{table}

\phead{Results.}
\noindent\textbf{\em Both repository-based review and the hierarchical evidence representation substantially contribute to repair performance.}
Table~\ref{tab:rq3_ablation_summary} summarizes the ablation results. The full \tool achieves a Pass@1 of 59.67\%. Removing repository-based review reduces Pass@1 to 48.00\%, corresponding to a 19.55\% relative decrease. Removing the hierarchical evidence representation reduces Pass@1 to 47.67\%, corresponding to a 20.11\% relative decrease. Both ablations remain above the original-report performance reported in RQ1 (41.00\%), indicating that repository exploration alone provides useful repair-relevant context. However, the substantial gap between the ablated variants and the full \tool shows that organizing hierarchical evidence and reviewing the generated specification are both important contributors to the overall repair gains achieved by \tool.

\rqboxc{
Ablation results on 300 SWE-Bench Lite instances show that removing repository-based review reduces Pass@1 from 59.67\% to 48.00\% ($\downarrow$19.55\%), while removing the hierarchical evidence representation reduces Pass@1 to 47.67\% ($\downarrow$20.11\%). Both components contribute substantially to the repair gains achieved by \tool.}

\subsection*{\RQFour}
\label{RQ4}

\begin{table*}[t]
\centering
\caption{Average per-instance computational cost of report generation and downstream repair on 300 SWE-Bench Lite instances. Report-generation costs are estimated from token usage and model pricing, while repair token usage and costs are recorded from Mini-SWE-Agent V2 execution logs. The original variant uses the benchmark bug report directly and therefore incurs no report-generation cost.}
\label{tab:rq4_cost}
\scalebox{0.85}{
\begin{tabular}{ll|rrr|rrr|rrr}
\toprule
\multirow{2}{*}{\textbf{Model}} 
& \multirow{2}{*}{\textbf{Report Variant}}
& \multicolumn{3}{c|}{\textbf{Report Generation}}
& \multicolumn{3}{c|}{\textbf{Repair}}
& \multicolumn{3}{c}{\textbf{Total}} \\
\cmidrule(lr){3-5} \cmidrule(lr){6-8} \cmidrule(lr){9-11}
& 
& \textbf{Input Tokens} 
& \textbf{Output Tokens} 
& \textbf{Cost}
& \textbf{Input Tokens} 
& \textbf{Output Tokens} 
& \textbf{Cost}
& \textbf{Input Tokens} 
& \textbf{Output Tokens} 
& \textbf{Cost} \\
\midrule
\multirow{3}{*}{GPT-5-mini}
& Original
& -- & -- & \$0.000
& 331K & 6K & \$0.027
& 331K & 6K & \$0.027 \\
& \baseline
& 85K & 5K & \$0.031
& 242K & 5K & \$0.021
& 327K & 10K & \$0.052 \\
& \tool
& 224K & 16K & \$0.087
& 251K & 5K & \$0.023
& 475K & 21K & \$0.110 \\
\midrule
\multirow{3}{*}{MiniMax M2.5}
& Original
& -- & -- & \$0.000
& 1{,}435K & 13K & \$0.075
& 1{,}435K & 13K & \$0.075 \\
& \baseline
& 50K & 2K & \$0.009
& 980K & 13K & \$0.056
& 1{,}030K & 15K & \$0.065 \\
& \tool
& 194K & 5K & \$0.034
& 1{,}089K & 12K & \$0.060
& 1{,}283K & 17K & \$0.094 \\
\bottomrule
\end{tabular}}
\end{table*}

\phead{Motivation.}
To generate refined reports, \tool explores the repository, organizes diagnostic evidence, generates a draft report, and applies repository-based review. These operations incur token and monetary costs. At the same time, the report provided to the repair agent may affect the cost of downstream repair. RQ4, therefore, evaluates both the cost of generating refined reports and the total cost of using those reports for automated program repair.

\phead{Approach.}
We measure two cost sources for each report variant. First, we measure \emph{report-generation cost}, which includes the tokens and estimated monetary cost needed to produce the \baseline and \tool reports. For generated report variants, this cost includes all pre-repair LLM calls, including trajectory collection, evidence extraction, hierarchical evidence abstraction, draft report generation, and repository-based review when applicable. The original report incurs no report-generation cost because it uses the benchmark bug reports directly. Second, we measure \emph{repair cost}, which includes the tokens and monetary costs incurred by Mini-SWE-Agent V2~\cite{minisweagentV2} during the repair process when given the corresponding report variant. For each variant, we report the report-generation cost, repair cost, and their sum as the total per-instance cost. All costs are reported as averages over the 300 SWE-Bench Lite~\cite{jimenez2024swebench} instances.

\phead{Results.} 
\noindent\textbf{\emph{\tool reduces downstream repair cost and input-token usage.}}
Table~\ref{tab:rq4_cost} reports the average per-instance costs of report generation, repair, and their combined total. With GPT-5-mini, \tool reports reduce downstream repair cost from \$0.027 for the original reports to \$0.023, while reducing repair input tokens from 331K to 251K. With MiniMax M2.5, \tool reports reduce downstream repair cost from \$0.075 for the original reports to \$0.060, while reducing repair input tokens from 1.435M to 1.089M. Thus, once the refined report is generated, \tool reduces repair input-token usage by approximately 24\% under both LLM backbones while also lowering the monetary cost of the downstream repair run.

\noindent\textbf{\emph{\tool achieves substantially higher repair success with a modest increase in absolute end-to-end cost.}}
When report generation and repair are considered together, \tool costs \$0.110 per instance with GPT-5-mini and \$0.094 with MiniMax M2.5. Although these totals are higher than those of the original and \baseline variants, they remain at most \$0.110 per instance. Compared with using the original reports, the additional cost is \$0.083 per instance with GPT-5-mini and \$0.019 with MiniMax M2.5. In return, as shown in RQ1 (Section~\ref{RQ1}), \tool reports improve Pass@1 by 45.53\% over the original reports and 22.60\% over \baseline reports with GPT-5-mini. With MiniMax M2.5, \tool reports improve Pass@1 by 17.68\% over the original reports and 21.38\% over \baseline reports. Overall, the refined reports incur a small absolute monetary overhead while producing substantially stronger repair outcomes. Once generated, they also reduce downstream repair cost and input-token consumption.

\rqboxc{
Including report generation and repair, \tool costs \$0.110 per instance with GPT-5-mini and \$0.094 with MiniMax M2.5. Although higher than using the original reports directly, the absolute increases are only \$0.083 and \$0.019 per instance, while Pass@1 improves by 45.53\% and 17.68\% for GPT-5-mini and MiniMax M2.5, respectively. Moreover, once the refined report is generated, downstream repair becomes cheaper.}

\section{Threats to Validity}
\label{sec:threats}

\noindent\textbf{Internal Validity.}
Our results depend on the correctness of trajectory collection, evidence abstraction, repository-based review, and downstream repair execution. Errors in repository checkout, tool execution, trajectory logging, or benchmark validation could affect the generated refined reports or measured Pass@1 outcomes. To mitigate this risk, we keep prompts, model settings, and refinement procedures fixed across instances. \tool uses only the original bug report and pre-fix repository during refinement, discards any candidate patch produced during the trajectory-collection run, and does not access the developer patch, post-fix code, or benchmark outcome. Since trajectories are unverified, they may contain weakly supported claims or abandoned hypotheses. \tool therefore treats trajectories only as candidate evidence and applies repository-based review before producing the final refined report. Although this prevents pipeline-level leakage, model-level memorization of public benchmark artifacts remains a possible threat in LLM-based studies.

\noindent\textbf{External Validity.}  
Our evaluation uses SWE-Bench Lite~\cite{jimenez2024swebench}, which consists of real-world GitHub issues from Python open-source projects. Although the design of \tool is language-agnostic, our evaluation does not establish its effectiveness for other programming languages or benchmarks. The results may also depend on the downstream repair agents and LLMs used in the evaluation. We partially mitigate this threat by evaluating \tool with Mini-SWE-Agent V2~\cite{minisweagentV2}, Agentless~\cite{xia2025agentless}, and AutoCodeRover~\cite{zhang2024autocoderover}, and with two LLM models, GPT-5-mini (\texttt{gpt-5-mini-2025-08-07})~\cite{gpt-5-mini} and  MiniMax M2.5 (\texttt{minimax-m2.5})~\cite{minimax-m2.5}. Further evaluation on additional benchmarks, ecosystems, and repair frameworks remains future work.

\noindent\textbf{Construct Validity.}  
We measure the effectiveness of refined reports using Pass@1 under the SWE-Bench Lite oracle~\cite{jimenez2024swebench}. This metric matches our goal of improving downstream APR, since an instance is counted as resolved only when the generated patch passes both the benchmark's fail-to-pass and pass-to-pass tests. However, Pass@1 measures end-to-end repair success rather than standalone properties of the refined report, such as readability, conciseness, or human-perceived diagnostic quality. Therefore, our conclusions focus on the utility of refined reports for automated repair rather than on all possible dimensions of report quality.

\section{Conclusion}
\label{sec:conclusion}

Bug reports serve as the primary task specifications for repository-level APR agents, but they often omit repair-relevant information needed to understand the failure, infer the intended behavior, and identify the implementation scope for repair. In this paper, we presented \tool, a trajectory-guided approach for repository-supported bug report specification refinement. \tool uses the trajectory produced by an unverified trajectory-collection run to recover specification evidence, organizes that evidence into a hierarchical representation, and applies repository-based review to produce a refined report for downstream repair. Our evaluation on all 300 SWE-Bench Lite instances shows that \tool improves Pass@1 from 41.00\% to 59.67\% with GPT-5-mini and from 54.67\% to 64.33\% with MiniMax M2.5, using Mini-SWE-Agent V2 as the downstream repair agent. On a repository-stratified 100-instance subset, \tool also improves Agentless from 41.00\% to 71.00\% and AutoCodeRover from 47.00\% to 72.00\%. These results show that trajectories can be reused beyond candidate patch generation: when abstracted, structured, and reviewed against the pre-fix repository, they provide actionable repository-supported evidence that improves the specifications guiding automated repair.

\balance
\bibliographystyle{IEEEtranN}
\bibliography{reference}

@article{chen2021demystifying,
  title={Demystifying the challenges and benefits of analyzing user-reported logs in bug reports},
  author={Chen, An Ran and Chen, Tse-Hsun Peter and Wang, Shaowei},
  journal={Empirical Software Engineering},
  volume={26},
  number={1},
  pages={1--30},
  year={2021},
  publisher={Springer}
}

@article{chen2021pathidea,
  title={Pathidea: Improving information retrieval-based bug localization by re-constructing execution paths using logs},
  author={Chen, An Ran and Chen, Tse-Hsun and Wang, Shaowei},
  journal={IEEE Transactions on Software Engineering},
  volume={48},
  number={8},
  pages={2905--2919},
  year={2021},
  publisher={IEEE}
}

@inproceedings{bettenburg2008makes,
  title={What makes a good bug report?},
  author={Bettenburg, Nicolas and Just, Sascha and Schr{\"o}ter, Adrian and Weiss, Cathrin and Premraj, Rahul and Zimmermann, Thomas},
  booktitle={Proceedings of the 16th ACM SIGSOFT International Symposium on Foundations of software engineering},
  pages={308--318},
  year={2008}
}

@inproceedings{rastkar2010summarizing,
  title={Summarizing software artifacts: a case study of bug reports},
  author={Rastkar, Sarah and Murphy, Gail C and Murray, Gabriel},
  booktitle={Proceedings of the 32nd ACM/IEEE International Conference on Software Engineering-Volume 1},
  pages={505--514},
  year={2010}
}

@misc{gpt-5-mini,
  title        = {GPT-5 mini Model},
  author       = {{OpenAI}},
  year         = {2025},
  howpublished ={\url{https://developers.openai.com/api/docs/models/gpt-5-mini}},
  note = {Accessed: 2026-06-26}
}

@misc{minimax-m2.5,
  author = {MiniMax},
  title = {MiniMax M2.5: Built for Real-World Productivity},
  year = {2026},
  howpublished ={\url{https://www.minimax.io/news/minimax-m25}},
  note = {Accessed: 2026-06-26}
}

@misc{minisweagentV2,
  author       = {Princeton NLP Group and Contributors},
  title        = {mini-SWE-Agent V2: The Minimal AI Software Engineering Agent},
  year         = {2025},
  howpublished = {\url{https://github.com/SWE-agent/mini-swe-agent}},
  note         = {Accessed: 2026-06-26}
}

@article{acharya2025can,
  title={Can We Enhance Bug Report Quality Using LLMs?: An Empirical Study of LLM-Based Bug Report Generation},
  author={Acharya, Jagrit and Ginde, Gouri},
  journal={arXiv preprint arXiv:2504.18804},
  year={2025}
}

@inproceedings{medeiros2024impact,
  title={The Impact Of Bug Localization Based on Crash Report Mining: A Developers' Perspective},
  author={Medeiros, Marcos and Kulesza, Uir{\'a} and Coelho, Roberta and Bonif{\'a}cio, Rodrigo and Treude, Christoph and Barbosa, Eiji Adachi},
  booktitle={Proceedings of the 46th International Conference on Software Engineering: Software Engineering in Practice},
  pages={13--24},
  year={2024}
}

@inproceedings{nayrolles2015jcharming,
  title={JCHARMING: A bug reproduction approach using crash traces and directed model checking},
  author={Nayrolles, Mathieu and Hamou-Lhadj, Abdelwahab and Tahar, Sofi{\`e}ne and Larsson, Alf},
  booktitle={2015 IEEE 22nd International Conference on Software Analysis, Evolution, and Reengineering (SANER)},
  pages={101--110},
  year={2015},
  organization={IEEE}
}

@article{yang2024swe,
  title={Swe-agent: Agent-computer interfaces enable automated software engineering},
  author={Yang, John and Jimenez, Carlos E and Wettig, Alexander and Lieret, Kilian and Yao, Shunyu and Narasimhan, Karthik and Press, Ofir},
  journal={Advances in Neural Information Processing Systems},
  volume={37},
  pages={50528--50652},
  year={2024}
}

@article{yu2025orcaloca,
  title={Orcaloca: An llm agent framework for software issue localization},
  author={Yu, Zhongming and Zhang, Hejia and Zhao, Yujie and Huang, Hanxian and Yao, Matrix and Ding, Ke and Zhao, Jishen},
  journal={arXiv preprint arXiv:2502.00350},
  year={2025}
}

@inproceedings{wang2025openhands,
  title={Openhands: An open platform for ai software developers as generalist agents},
  author={Wang, Xingyao and Li, Boxuan and Song, Yufan and Xu, Frank F and Tang, Xiangru and Zhuge, Mingchen and Pan, Jiayi and Song, Yueqi and Li, Bowen and Singh, Jaskirat and others},
  booktitle={International Conference on Learning Representations},
  volume={2025},
  pages={65882--65919},
  year={2025}
}

@article{xia2025live,
  title={Live-SWE-agent: Can Software Engineering Agents Self-Evolve on the Fly?},
  author={Xia, Chunqiu Steven and Wang, Zhe and Yang, Yan and Wei, Yuxiang and Zhang, Lingming},
  journal={arXiv preprint arXiv:2511.13646},
  year={2025}
}

@inproceedings{dang2012rebucket,
  title={Rebucket: A method for clustering duplicate crash reports based on call stack similarity},
  author={Dang, Yingnong and Wu, Rongxin and Zhang, Hongyu and Zhang, Dongmei and Nobel, Peter},
  booktitle={2012 34th International Conference on Software Engineering (ICSE)},
  pages={1084--1093},
  year={2012},
  organization={IEEE}
}

@inproceedings{jimenez2024swebench,
  title={Swe-bench: Can language models resolve real-world github issues?},
  author={Jimenez, Carlos E and Yang, John and Wettig, Alexander and Yao, Shunyu and Pei, Kexin and Press, Ofir and Narasimhan, Karthik},
  booktitle={International Conference on Learning Representations},
  volume={2024},
  pages={54107--54157},
  year={2024}
}

@inproceedings{xia2025agentless,
  title={Demystifying llm-based software engineering agents},
  author={Xia, Chunqiu Steven and Deng, Yinlin and Dunn, Soren and Zhang, Lingming},
  journal={Proceedings of the ACM on Software Engineering},
  volume={2},
  number={FSE},
  pages={801--824},
  year={2025},
}

@inproceedings{zhang2024autocoderover,
author = {Zhang, Yuntong and Ruan, Haifeng and Fan, Zhiyu and Roychoudhury, Abhik},
title = {AutoCodeRover: Autonomous Program Improvement},
year = {2024},
isbn = {9798400706127},
publisher = {Association for Computing Machinery},
booktitle = {Proceedings of the 33rd ACM SIGSOFT International Symposium on Software Testing and Analysis},
pages = {1592–1604}
}

@inproceedings{agent_traj_thought_action_result,
author = {Bouzenia, Islem and Pradel, Michael},
title = {Understanding Software Engineering Agents: A Study of Thought-Action-Result Trajectories},
booktitle = {2025 40th IEEE/ACM International Conference on Automated Software Engineering (ASE)},
pages = {2846–2857},
numpages = {12},
year = {2025}
}

@article{soltani2020significance,
  title={The significance of bug report elements},
  author={Soltani, Mozhan and Hermans, Felienne and B{\"a}ck, Thomas},
  journal={Empirical Software Engineering},
  volume={25},
  number={6},
  pages={5255--5294},
  year={2020},
  publisher={Springer}
}

@inproceedings{hirsch2020root,
  title={Root cause prediction based on bug reports},
  author={Hirsch, Thomas and Hofer, Birgit},
  booktitle={2020 IEEE International Symposium on Software Reliability Engineering Workshops (ISSREW)},
  pages={171--176},
  year={2020},
  organization={IEEE}
}

@article{fahim2025crash,
  title={Crash Report Enhancement with Large Language Models: An Empirical Study},
  author={Fahim, SM and Rafi, Md Nakhla and Ma, Zeyang and Kim, Dong Jae and others},
  journal={arXiv preprint arXiv:2509.13535},
  year={2025}
}

@inproceedings{bouzenia2025repairagent,
  title={Repairagent: An autonomous, llm-based agent for program repair},
  author={Bouzenia, Islem and Devanbu, Premkumar and Pradel, Michael},
  booktitle={2025 IEEE/ACM 47th International Conference on Software Engineering (ICSE)},
  pages={2188--2200},
  year={2025},
  organization={IEEE}
}

@inproceedings{al2025llput,
  title={Llput: Investigating large language models for bug report-based input generation},
  author={Al Hasan, Alif and Saha, Subarna and Imran, Mia Mohammad and Zaman, Tarannum Shaila},
  booktitle={Proceedings of the 33rd ACM International Conference on the Foundations of Software Engineering},
  pages={1652--1659},
  year={2025}
}

@inproceedings{suri2026codescout,
  title={Codescout: Contextual problem statement enhancement for software agents},
  author={Suri, Manan and Li, Xiangci and Shojaie, Mehdi and Han, Songyang and Hsu, Chao-Chun and Garg, Shweta and Deshmukh, Aniket Anand and Kumar, Varun},
  booktitle={Findings of the Association for Computational Linguistics: ACL 2026},
  pages={40902--40931},
  year={2026}
}

@article{kuang2026reagent,
  title={REAgent: Requirement-Driven LLM Agents for Software Issue Resolution},
  author={Kuang, Shiqi and Tian, Zhao and Lin, Kaiwei and Tao, Chaofan and Wang, Shaowei and Bai, Haoli and Shang, Lifeng and Chen, Junjie},
  journal={arXiv preprint arXiv:2604.06861},
  year={2026}
}

@article{tamoyan2026sherloc,
  title={SHERLOC: Structured Diagnostic Localization for Code Repair Agents},
  author={Tamoyan, Hovhannes and Narenthiran, Sean and Arakelyan, Erik and Mezini, Mira and Ginsburg, Boris},
  journal={arXiv preprint arXiv:2606.24820},
  year={2026}
}

@article{zhang2026sgagent,
  title={Sgagent: Suggestion-guided llm-based multi-agent framework for repository-level software repair},
  author={Zhang, Quanjun and Gao, Chengyu and Han, Yu and Shang, Ye and Fang, Chunrong and Chen, Zhenyu and Xiao, Liang},
  journal={ACM Transactions on Software Engineering and Methodology},
  year={2026},
  publisher={ACM New York, NY}
}

@article{yang2025enhancing,
  title={Enhancing repository-level software repair via repository-aware knowledge graphs},
  author={Yang, Boyang and Ren, Jiadong and Jin, Shunfu and Liu, Yang and Liu, Feng and Le, Bach and Tian, Haoye},
  journal={arXiv preprint arXiv:2503.21710},
  year={2025}
}

@article{pan2026reporepair,
  title={RepoRepair: Leveraging Code Documentation for Repository-Level Automated Program Repair},
  author={Pan, Zhongqiang and Li, Chuanyi and Zhong, Wenkang and Feng, Yi and Luo, Bin and Ng, Vincent},
  journal={arXiv preprint arXiv:2603.01048},
  year={2026}
}

@article{mu2025experepair,
  title={Experepair: Dual-memory enhanced llm-based repository-level program repair},
  author={Mu, Fangwen and Wang, Junjie and Shi, Lin and Wang, Song and Li, Shoubin and Wang, Qing},
  journal={arXiv preprint arXiv:2506.10484},
  year={2025}
}

@article{li2026outcome,
  title={Outcome-Conditioned Reasoning Distillation for Resolving Software Issues},
  author={Li, Chenglin and Xu, Yisen and Wang, Zehao and Tan, Shin Hwei and others},
  journal={arXiv preprint arXiv:2601.23257},
  year={2026}
}

@article{zhu2026swe,
  title={Swe context bench: A benchmark for context learning in coding},
  author={Zhu, Jiayuan and Wu, Junde and Hu, Minhao and Zhu, Shengda and Pan, Jiazhen and Shen, Weixiang and Yang, Yijun and Liu, Fenglin and Hao, Jianye and Jin, Yueming and others},
  journal={arXiv preprint arXiv:2602.08316},
  year={2026}
}

@article{pei2025scope,
  title={Scope: Prompt evolution for enhancing agent effectiveness},
  author={Pei, Zehua and Zhen, Hui-Ling and Kai, Shixiong and Pan, Sinno Jialin and Wang, Yunhe and Yuan, Mingxuan and Yu, Bei},
  journal={arXiv preprint arXiv:2512.15374},
  year={2025}
}

@article{zhang2026coda,
  title={CODA-BENCH: Can Code Agents Handle Data-Intensive Tasks?},
  author={Zhang, Yuxin and Fan, Ju and Fan, Meihao and Zhang, Shaolei and Du, Xiaoyong},
  journal={arXiv preprint arXiv:2606.15300},
  year={2026}
}

@article{lyu2026agentszz,
  title={AgentSZZ: Teaching the LLM Agent to Play Detective with Bug-Inducing Commits},
  author={Lyu, Yunbo and Shi, Jieke and Kang, Hong Jin and Widyasari, Ratnadira and He, Junda and Niu, Yuqing and Yang, Chengran and Chen, Junkai and Yang, Zhou and Lawall, Julia and others},
  journal={arXiv preprint arXiv:2604.02665},
  year={2026}
}

@article{yang2026twinrouterbench,
  title={TwinRouterBench: Fast Static and Live Dynamic Evaluation for Realistic Agentic LLM Routing},
  author={Yang, Pei and Chen, Wanyi and Yang, Tongyun and Feng, Pengbin and Xing, Jiarong and Guo, Wentao and Yao, Yuhang and Han, Yuhang and Li, Hanchen and Wang, Xu and others},
  journal={arXiv preprint arXiv:2605.18859},
  year={2026}
}

@article{li2026dualeval,
  title={DualEval: Joint Model-Item Calibration for Unified LLM Evaluation},
  author={Li, Aaron J and Huang, Hao and Park, Youngmin and Ma, Yitong and Chiang, Wei-Lin and Chen, Li and Hsieh, Cho-Jui and Yu, Bin and Stoica, Ion},
  journal={arXiv preprint arXiv:2606.26429},
  year={2026}
}

@article{joshi2026inference,
  title={Inference Time Context Sparsity: Illusion or Opportunity?},
  author={Joshi, Sahil and Dixit, Prithvi and Chowdhury, Agniva and Shrivastava, Anshumali and Gonzalez, Joseph E and Stoica, Ion and Agrawal, Kumar Krishna and Desai, Aditya},
  journal={arXiv preprint arXiv:2605.24168},
  year={2026}
}

@article{zhu2026specification,
  title={Specification Vibing for Automated Program Repair},
  author={Zhu, Taohong and Cordeiro, Lucas C and Mustafa, Mustafa A and Sun, Youcheng},
  journal={arXiv preprint arXiv:2602.08263},
  year={2026}
}

@article{huang2025dynafix,
  title={DynaFix: Iterative Automated Program Repair Driven by Execution-Level Dynamic Information},
  author={Huang, Zhili and Xu, Ling and Liu, Chao and Sun, Weifeng and Zhang, Xu and Lei, Yan and Yan, Meng and Zhang, Hongyu},
  journal={arXiv preprint arXiv:2512.24635},
  year={2025}
}

@article{akyol2026improbr,
  title={ImproBR: Bug Report Improver Using LLMs},
  author={Akyol, Emre Furkan and Dedeler, Mehmet and T{\"u}z{\"u}n, Eray},
  journal={arXiv preprint arXiv:2604.26142},
  year={2026}
}

\end{document}